\documentclass[12pt]{elsarticle}
\usepackage{lineno}
\usepackage{epsfig}
\usepackage{amssymb}
\usepackage{setspace}

\journal{Nucl. Instrum. Meth. A}

\begin{document}
\begin{frontmatter}

\title{Response of CdWO$_4$ crystal scintillator for few MeV ions and low energy electrons}

\author[DP-INFN-Fi] {P.G.~Bizzeti}
\author[DP-INFN-Fi] {L.~Carraresi}
\author[KINR]       {F.A.~Danevich}
\author[DP-INFN-Fi] {T.~Fazzini}
\author[DP-INFN-Fi] {P.R.~Maurenzig}
\author[DP-INFN-Fi] {F.~Taccetti}
\author[DP-INFN-Fi] {N.~Taccetti}
\author[KINR]       {V.I.~Tretyak}



\address[DP-INFN-Fi]{Dipartimento di Fisica, Universita di Firenze and INFN sez. Firenze, I-50019, Italy}
\address[KINR]{Institute for Nuclear Research, MSP 03680 Kyiv, Ukraine}

\begin{abstract}


The response of a CdWO$_4$ crystal scintillator to protons, $\alpha$ particles, 
Li, C, O and Ti ions with energies in the range 1 -- 10 MeV was 
measured. The non-proportionality of CdWO$_4$ for 
low energy electrons (4 -- 110 keV) was studied with the Compton Coincidence
Technique. The energy dependence of the quenching factors for ions and
the relative light yield for low energy electrons was
calculated using a semi-empirical approach. Pulse-shape
discrimination ability between $\gamma$ quanta, protons,
$\alpha$ particles and ions was investigated.

\end{abstract}

\begin{keyword}

Scintillation detectors \sep CdWO$_4$ crystal \sep Quenching \sep
Non-proportionality in scintillation response \sep Scintillation
pulse-shape \sep Dark matter detection

\PACS 23.40.-s

\end{keyword}
\end{frontmatter}

\section{Introduction}

Numerous astronomical and cosmological observations since 1930's
suggest that most of the matter in the Universe is non-luminous
and non-baryonic, while usual matter constitutes only $\simeq4\%$ of 
the Universe and the main components are dark matter
($\simeq23\%$) and dark energy ($\simeq73\%$) \cite{Feng10,Bert10}. 
Dark matter (DM) is preferably related with particles which are neutral 
and only weakly interacting with matter (Weakly Interacting Massive Particles, 
WIMPs). Interaction of such particles with usual matter could be 
detected through observation of nuclear recoils created 
after scattering of WIMPs on atomic nuclei in sensitive detectors
placed in low background conditions deep underground.
Positive evidence of WIMPs observation at $\simeq 9 \sigma$ confidence 
level is reported in the DAMA experiment \cite{Bern08} after 13 
years measurements with large mass low background NaI(Tl) scintillators.
This result was recently supported by the CoGeNT \cite{Aal11} 
and CRESST \cite{Ang12} data. However, many other 
searches for WIMPs to-date gave only negative results \cite{Spoo07,Rau11}.
Because of the fundamental importance of the question of the DM
constituents, searches for WIMPs are under performance or planned
in near twenty experiments, in particular, in the ton-scale
projects EURECA with cryogenic Ge and 
different crystal scintillating bolometers 
\cite{Krau07}, DARWIN with scintillating noble gases \cite{Baud12}
and DAMA/1ton (proposed since 1996) \cite{Cape07}.

When the energy released by the recoil ions is measured with scintillators,
one of the main questions is the quenching of the scintillation light
yield. Since a long time it is known that the amount of light produced
in scintillating materials 
by highly ionizing particles is lower
than that produced by electrons of the same energy \cite{Birk64}.
In a scintillator calibrated with $\gamma$ sources,
signals from ions will be seen at lower energies than their real values,
sometimes by more than one order of magnitude. 
Knowledge of quenching factor, QF (i.e. ratio of the measured ion energy in 
$\gamma$ scale to its real energy) is very important in searches for 
WIMPs and in predictions where the WIMPs signal should be expected. 
QFs depend on many factors such as:
the scintillating material itself, its dopants and impurities,
temperature; ion's $Z$ and $A$ numbers, ion's energy; time of
collection of scintillating signal, etc. (see some examples in
\cite{Tret10}). Since it is sometimes  quite difficult to measure
QFs for the needed ions and in the needed energy region (f.e. for
low-energy heavy ions when the scintillation signal is expected at 
$\simeq10$ keV or less), some methods for the QFs estimation are 
of great interest.

QFs for different ions in a scintillator could
be not independent but related quantities. Such a hypothesis,
supported by some experimental data, was discussed already in
\cite{Birk64}.  
If true, on the basis of measurements of QFs for
particles of one kind in some energy region (e.g. for a few
MeV $\alpha$ particles from internal contamination of a detector), one
would be able to {\it calculate} QFs for particles of another kind
and for other energies (e.g. for low energy nuclear recoils).
Further evidences in favor of this hypothesis were 
given in \cite{Tret10}.

Cadmium tungstate (CdWO$_4$) crystal scintillators are widely used
in low counting experiments to search for $2\beta$ decay
\cite{Dan96a,Dan03a,Bel08,Bar11,Bel12} and studies of rare $\beta$
\cite{Dan96b,Bel07} and $\alpha$ \cite{Dan03b} decays. CdWO$_4$ has
rather similar properties to CaWO$_4$, ZnWO$_4$, CaMoO$_4$ and 
some other oxide crystal scintillators, promising targets for DM
experiments \cite{Krau07}.

In this work we measured the response of a CdWO$_4$ crystal 
to few MeV energy ions and low energy ($4-110$ keV) electrons. 
Pulse-shape discrimination ability was studied
for protons, $\alpha$ particles, Li, C, O and Ti ions. 
Using the measured quenching for protons, we applied a 
semi-empirical approach \cite{Tret10} to 
calculate QFs for other ions and low energy electrons.

\section{\label{s:DUS}Crystal characteristics and signal processing}

The CdWO$_4$ crystal studied in this work is a parallelepiped of
$10\times20\times25$ mm, manufactured by Scionix. CdWO$_4$
is a monoclinic, almost orthorhombic, crystal. From the
specification of the manufacturer, the $10\times20$ mm faces are
known to be cleavage (010) planes. Measurements performed with a
diffractometer\footnote{These measurements have been performed at
the ``Centro di Cristallografia Strutturale'' of the University of
Firenze.} confirm this assignment, while the two faces
$10\times25$ mm and $20\times25$ mm almost correspond to planes
of indexes (100) and (001), respectively.

The crystal is optically coupled through a $10\times20$ mm face
to a photomultiplier (PMT) ETL mod. 9256B. Its photocathode has an
extended green response for a better matching with the light
emission spectrum of the CdWO$_4$ crystal. In the measurements with
low energy electrons the crystal has been wrapped with Teflon
tape, while in measurements under beam an aperture of
$10\times10$ mm has been opened in the tape  covering one $20\times25$ mm
face, so that ions can enter directly. It has been verified with
$\gamma$ sources that no measurable difference in light collection
is found with or without aperture. 
The energy scale of the detector was established with $^{22}$Na,
$^{57}$Co, $^{60}$Co, $^{137}$Cs and $^{241}$Am calibration sources.

In all the measurements the anode signals from the PMT are
processed by a current to voltage converter which acts also as
antialiasing filter and analysed by a transient
digitiser\footnote{The main characteristics are: 12 bits
(11-effective), 20 Ms/s, $\pm 1024$ mV linear range.} 
already described in \cite{Bard06}. The data, sequentially
digitised every 50 ns, are stored in a temporary memory FIFO,
which, in the presence of an event trigger, is stopped and read in a time
interval starting about 30 $\mu$s before the time of the event and
extending  to 128 $\mu$s on the whole. These data are stored on a
mass memory for further analysis. In particular the amplitude
spectra are obtained by the following procedure: for each event
the mean value of the baseline  is evaluated in the first 25 $\mu$s 
and subtracted channel by channel, afterwards the signal
amplitude is obtained by summing up the channel contents in a
time interval lasting 55 $\mu$s from the beginning of the signal.
It is also checked whether the analysed waveform is in true or
chance coincidence with the event trigger; overlapping waveforms
are recognised and discarded.

\section{Response to protons, $\alpha$ particles, Li, C, O and Ti ions}

\subsection{Measurements with pulsed beams}

Beams of protons, Li, C, O, and Ti ions, in the energy range $1-10$ MeV, were produced 
by the Tandetron accelerator of LABEC at the INFN-Florence. 
A dedicated beam line \cite{Tacc05} provides short pulses by means of an electrostatic deflector
which displaces the beam spot across a narrow window obtained by an adjustable slits 
system, housed at the entrance of the vacuum chamber, situated at the end of the beam-line 
and containing the crystal. 
The beam intensity and the width of the slits were adjusted to obtain the arrival 
of a single particle per pulse in a large majority of the cases. As already said 
in Section 2, in all these measurements the accelerated particles enter the crystal 
perpendicularly to a 20 mm x 25 mm (001) face.
In these measurements the event trigger was derived from the voltage transition of 
the deflecting plates. Owing to this coincidence constraint on the stored pulse shapes,
it is possible to build from them energy spectra almost free from background.
The measured quenching factors  for protons (1.0, 2.0, 3.0, 4.0 MeV), Li (3.0, 4.5 MeV),
C (3.0, 5.1, 7.5 MeV), O (6.0 MeV) and Ti (10.0 MeV) ions are presented in Fig.~1.

\nopagebreak
\begin{figure}[h]
\begin{center}
\epsfig{figure=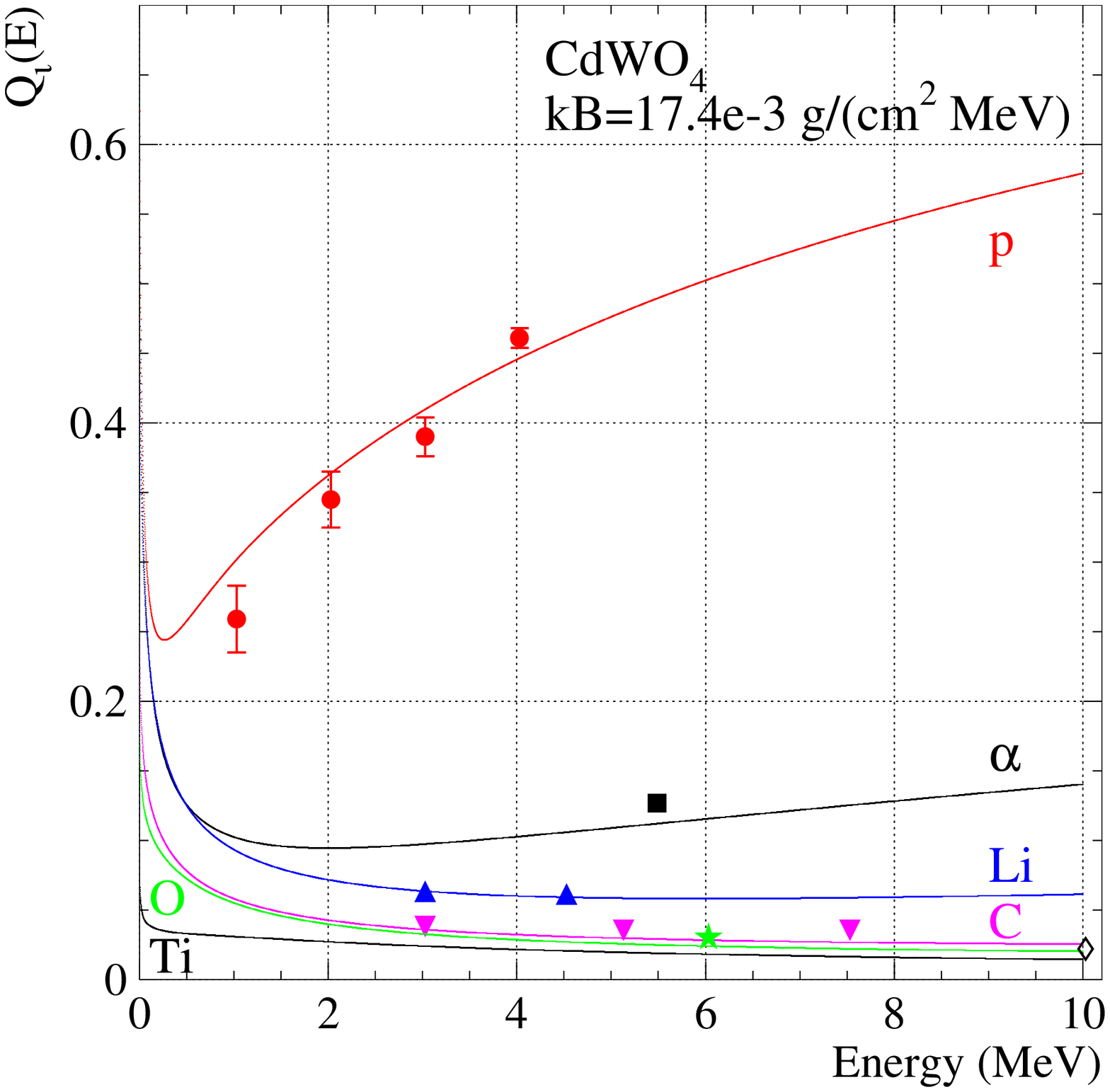,height=6.5cm}~
\epsfig{figure=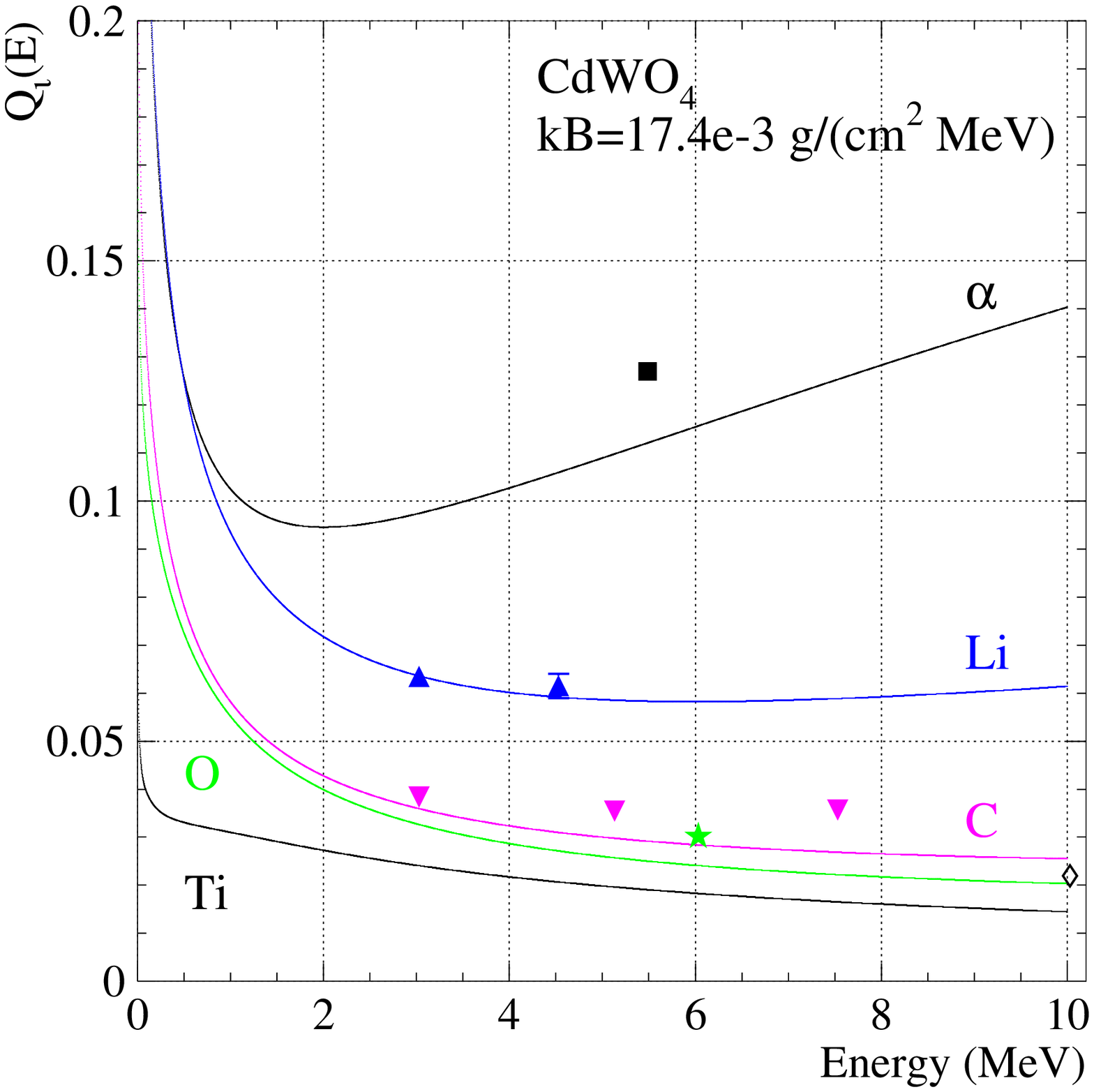,height=6.5cm}
\caption{(Color online) Quenching factors for
protons ($\bullet$), 
$\alpha$ particles ($\blacksquare$), 
Li ($\blacktriangle$), 
C ($\blacktriangledown$), 
O ($\bigstar$) and 
Ti ($\lozenge$) ions
measured with the CdWO$_4$ crystal scintillator. Solid lines
represent calculations in accordance with \cite{Tret10}.
On the right, lower QFs are shown in more detail.}
\end{center}
\end{figure}

Differences of the pulse shapes, as a consequence of different specific energy losses,
have been also investigated.

\subsection{Measurements with $\alpha$ particles}

In addition to measurements with accelerated ions, the CdWO$_4$ crystal was irradiated with
$\alpha$ particles. A collimated $^{241}$Am $\alpha$ source was placed, 
inside the crystal chamber, 
in front of the crystal (001) face and, at the same time, the crystal was irradiated 
by $^{137}$Cs and $^{60}$Co $\gamma$ sources in order to get a reference
$\gamma$ energy scale. In this 
case a self-trigger was used and, taking advantage from the different zero-crossing 
times,  $\alpha$ particle and  $\gamma$ ray spectra were easily disentangled.
From this measurement the quenching factor for the 5486 keV  $\alpha$ particles, entering 
perpendicularly the (001) face, was found to be 0.127(1) as reported in Fig.~1.
In a different series of measurements, the three faces of the bare crystal have 
been irradiated with  $\alpha$ particles. The measured QFs were: 0.197(1) for (010) face, 
0.167(1) for (100) face and 0.134(1) for (001) face. These measurements can be 
helpful for determining the response of the crystal to particles created in its 
volume and moving in random directions with respect to the principal planes.
A comment is required by the small ($\leq5\%$) but significant difference between the 
two QF values relative to the (001) face: most probably, the difference could be 
attributed to the different light losses associated to the different locations 
of  $\alpha$ and $\gamma$ light emitting sources (a surface point-like light source in the case 
of  $\alpha$ particles and a volume distributed light source in the case of $\gamma$ rays) with 
respect to the different reflecting and diffusing properties of the bare and Teflon 
wrapped crystal.

Difference in QFs for $\alpha$ particles moving in different directions, in addition to
CdWO$_4$ detectors (here and e.g. in \cite{Dan03b}) was observed also for ZnWO$_4$ \cite{Dan05}
and MgWO$_4$ \cite{Dan09} crystal scintillators. 
As quenching depends on ionization density \cite{Birk64}, such a behavior of the QFs 
has to be related with difference in ionization density and stopping power
for $\alpha$ particles moving in different directions inside an anisotropic crystal. 
This interpretation is also supported by the behavior of the shape indicator (Section 3.4),
which also depends on the direction of the $\alpha$ particles, being closer to that of 
$\gamma$s when the quenching factor is larger. Differences of light 
collection due to surface effects could, in fact, change the quenching 
factor but hardly modify the shape of the light pulse.

\subsection{Calculations of quenching factors for ions}

Quenching factors $Q_i$ for protons, $\alpha$ particles and Li, C, O, Ti 
ions in the CdWO$_4$ scintillator were calculated following the work
\cite{Tret10}, which is based on the classical Birks formula \cite{Birk64},
as the ratio of light yield of an ion to that of an electron of the same
energy:

\begin{equation}
Q_i(E) = L_i(E) / L_e(E),
\end{equation}

\noindent where  

\begin{equation}
L_i(E) = \int_0^E dL_i = \int_0^E \frac{S_idE}{1+kB(\frac{dE}{dr})_i},
\end{equation}

\begin{equation}
L_e(E) = \int_0^E dL_e = \int_0^E \frac{S_edE}{1+kB(\frac{dE}{dr})_e}.
\end{equation}

Here $(dE/dr)_i$ and $(dE/dr)_e$ are the total stopping powers for ions 
calculated with the SRIM code \cite{SRIM} and electrons calculated with the 
ESTAR code \cite{ESTAR}, respectively.

Supposing the normalization factors $S_{i,e}$ equal for electrons and ions and 
independent on energy, we obtain that $Q_i(E)$ depends only on a single parameter 
$kB$ (the Birks factor). It is assumed that $kB$ is also independent of energy and
has the same value for all ions, if all data are measured under the same
experimental conditions and treated in the same way \cite{Tret10}.

To determine the $kB$ value for the present CdWO$_4$ measurements, 
the experimental points for protons were fitted by curve calculated 
with Eq.~(1)--(3). 
The result is $kB=17.4$ mg~cm$^{-2}$~MeV$^{-1}$, and the obtained curve 
is shown in Fig.~1. 
Afterwards, QFs for all other ions were calculated with this $kB$ value;
all the results are also demonstrated in Fig.~1. 

As one can see, the calculations do not perfectly reproduce the 
experimental points, with the biggest deviation ($\simeq30\%$) 
for Ti ions. Nevertheless, general agreement could be considered
as acceptable, especially for a theory with only one parameter.
Sometimes quenching factors in DM experiments are known with much 
bigger uncertainties, and calculations in accordance with 
the above described approach could give valuable information on 
the expected QF values.

\subsection{Pulse-shape discrimination between $\gamma$ quanta, 
protons, $\alpha$ particles and light ions}

To test the ability of pulse-shape discrimination between $\gamma$
quanta ($\beta$ particles), protons, $\alpha$ particles and the light
ions, 
the data were analyzed by
using the optimal filter method proposed by E.~Gatti and F.~De
Martini \cite{Gat62} (see also \cite{Faz98} where the analysis was
developed for CdWO$_4$ crystal scintillators). For each
experimental signal,  its shape
indicator  was defined as $SI=\sum f(t_k)\times
P(t_k)/\sum f(t_k)$, where the sum is over time channels $k$,
starting from the origin of signal and up to 60 $\mu$s, $f(t_k)$
is the digitized amplitude of a given signal.
The weight function $P(t)$ is defined as:
$P(t) = \overline{f}_\alpha(t)-\overline{f}_\gamma(t)$, where
the reference pulse shapes $\overline{f}_\alpha(t)$ and
$\overline{f}_\gamma(t)$ are the average of a few thousands shapes for
$\alpha$ particles and $\gamma$ quanta, respectively, collected in
the measurements with $\gamma$ and $\alpha$ sources.

Distributions of the shape indicator versus energy for $\gamma$
quanta, protons, $\alpha$ particles, Li, C, O and Ti ions measured
with CdWO$_4$ crystal scintillator are presented in Fig.~2. 
There is clear discrimination between $\gamma$s, protons, $\alpha$
particles and ions. 
The shape indicator values for protons and ions 
lie near the ones of $\alpha$ particles. 
This fact, however, does not necessarily 
imply that the pulse shape be equal for heavier ions and protons. 

\nopagebreak
\begin{figure}[h]
\begin{center}
\mbox{\epsfig{figure=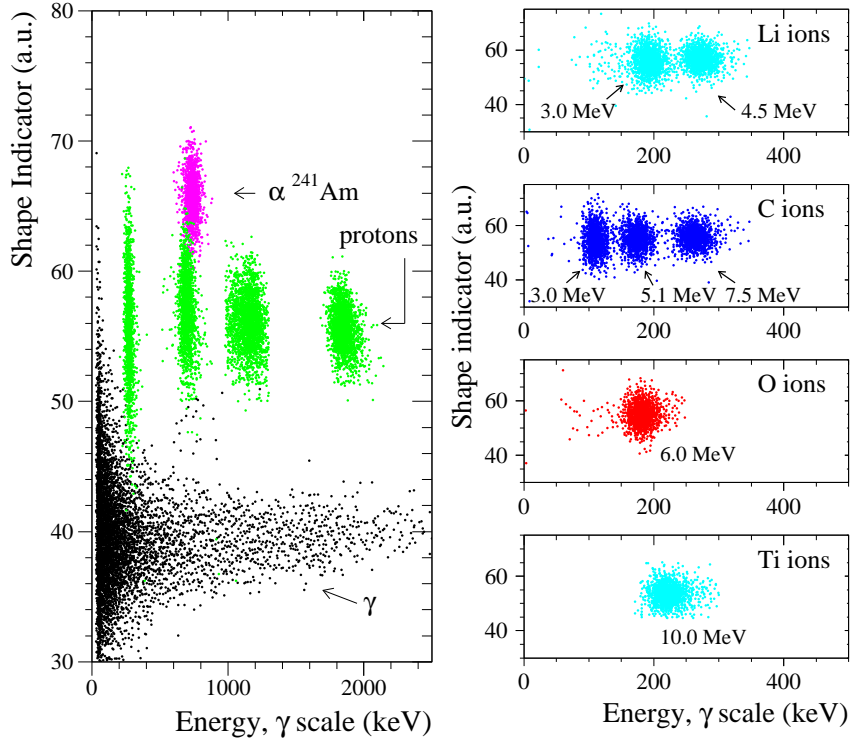,height=10.0cm}} 
\caption{(Color online) 
Distributions of shape indicator (see text) versus energy
for protons, Li, C, O and Ti ions measured with 
the CdWO$_4$ crystal
scintillator at the accelerator. Data for $\gamma$
quanta were obtained from a background run (mainly environmental
radioactivity), while $\alpha$ particles were accumulated with
the $^{241}$Am source installed in the vacuum chamber of the
accelerator.}
\end{center}
\end{figure}

It is worth to note also that the measurements with $^{241}$Am $\alpha$
particles entering into different faces of the CdWO$_4$ crystal 
(Section 3.2) show a clear relation between quenching factor and 
shape indicator: for bigger $QF_\alpha$, when $\alpha$ signal is 
closer to signal caused by $\gamma$ quanta, $SI_\alpha$ value is
lower, i.e. also closer to $SI_\gamma$. This effect was already observed 
in \cite{Dan03b,Dan05}.

\section{Measurements with low energy electrons}

It should be noted that the quenching of scintillation signals is
observed not only for ions but also for low-energy (less than a
few hundred keV) $\gamma$ quanta and electrons. However, the
nature of the electron-induced light-emitting excitation can have a
different origin from that induced by heavy ions. Possible relation
between QFs for ions and QFs for electrons is interesting and
important as giving an additional method to obtain QFs for recoils
induced by WIMPs in scintillators.

A lot of work has been dedicated, also recently, to the study of
the non-linearity of the response of scintillating crystals,
coupled to different types of photo-detectors (photomultipliers in
many cases), to low-energy $\gamma$ quanta and electrons (see e.g.
\cite{Saka87,Dore95,Jaff07,Chon08,Pay09,Mos10,Pay11,Kho12} 
and works quoted therein). 
We have studied the response of a CdWO$_4$ scintillation
detector to low energy electrons by using electrons created in
the CdWO$_4$ crystal by Compton scattered $\gamma$ quanta.

\subsection{Experimental set-up}

To avoid surface effects for low energy electrons, the CdWO$_4$
crystal under study is used as a Compton
spectrometer\footnote{This method was first proposed by  Valentine
and Rooney \cite{Vale94,Roon96}.}. This technique requires that
the $\gamma$ rays, scattered by the crystal at an angle $\theta$ with
respect to the direction of the incoming collimated monochromatic
beam, be detected in coincidence with the signals generated in
the crystal by the corresponding scattered electrons. However,
when the energy of the scattered electrons is small, typically
less than 20 keV, it becomes difficult to achieve reliable
information both in energy and in timing, particularly for
crystals, such as CdWO$_4$, whose main light emission is
characterised by a long decay time ($\approx 14~\mu$s)
\cite{Bard08}: in that case a small anode signal consists of a
sequence of almost randomly spaced single electron responses.

To overcome this problem, we have used the production in opposite
directions of the two  511 keV $\gamma$ rays from the singlet
positronium following the $\beta^+$ decay of $^{22}$Na. This
variant of the Compton Coincidence Technique makes it possible to
obtain a good timing of the event trigger through the coincidence
between one of the $\gamma$ rays observed by a  NaI(Tl) counter and
the other one observed in a high purity germanium (HPGe) counter, after been scattered in
the crystal under study, and therefore to study the response to
electrons down to very low energy also for CdWO$_4$. Moreover the
collimation of the $\gamma$ rays, impinging on the crystal, is
performed {\it via} coincidences with NaI(Tl), so avoiding the
border effects of a physical collimator. The front-end electronics
for the NaI(Tl) detector consists of a passive integrator of the anode
current signals followed by a JFET  input linear buffer feeding a
semi-gaussian shaping amplifier (Ortec  mod. 572: time constant
0.5 $\mu$s) whose bipolar output is sent to a Timing Single-Channel
Analyzer (Ortec, mod. 551). HPGe signals, from one of the twin
outputs of the charge preamplifier, are processed in the same way
as those of the NaI(Tl) detector and provide timing information,
while the other preamplifier output is connected to a Gated
Integrator Amplifier (Ortec: mod. 973) with integration time
5 $\mu$s, and used for energy measurement.

The geometry of the experimental set-up is shown in Fig.~3. 
The $^{22}$Na source has nominal activity 400 kBq. 
$^{22}$Na decays to the 1274.5 keV, 
first excited level of  $^{22}$Ne, through a $\beta^+$
($E_{max}$ = 546 keV) annihilating in a 2~mm diameter zone.

\nopagebreak
\begin{figure}[h]
\label{fig:set-up}
\begin{center}
\mbox{\epsfig{figure=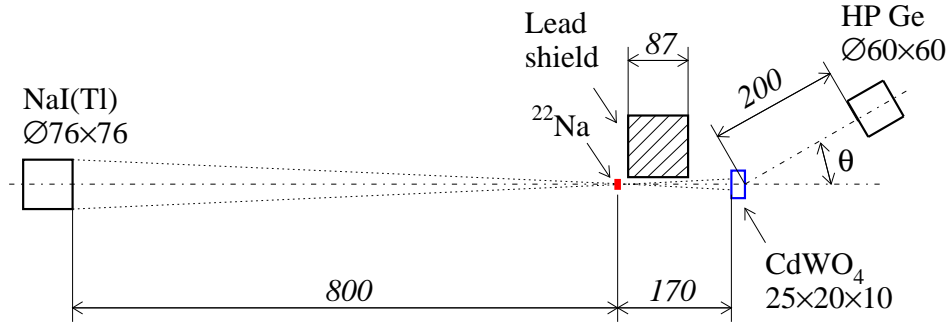,height=4.7cm}} \caption{(Color
online) Set-up for measurements of CdWO$_4$ crystal scintillator
response to low energy electrons. All sizes are in mm. The angle
$\theta$ has been set to 12.5, 20 and 35$^\circ$. The Al cap
surrounding the CdWO$_4$ crystal is not shown.}
\end{center}
\end{figure}

As shown in Fig.~3, the source is on the axis of a system of two
detectors: a $3^{''}\times 3^{''}$ NaI(Tl) (manufactured by
Scionix) and the CdWO$_4$ crystal described in Section
\ref{s:DUS}. The 10 mm thickness of this crystal causes an
attenuation of the 511 keV $\gamma$ rays of $\approx$60\%. Special
care has been devoted to minimise the material surrounding the
CdWO$_4$ crystal and thus the probability of absorption and
scattering along the path of the 511 keV $\gamma$ rays: to this
purpose the cap surrounding the CdWO$_4$ is a cylinder of
aluminium of only 0.4 mm thickness. The 800 mm distance between
source and front face of the NaI(Tl) detector defines a solid
angle of maximum angular aperture of 5.4$^\circ$, so that a 511
keV $\gamma$ ray impinging on this detector has a companion 511
keV $\gamma$ ray hitting the CdWO$_4$ crystal inside a circle of
16 mm diameter.

The 511 keV  $\gamma$ rays scattered at angle $\theta$ from the
crystal are detected by an HPGe detector (Ortec: mod. GMX 30 P:
diameter 60 mm, height 60 mm, relative efficiency 30\% and
resolution (FWHM) 1.9 keV at 1.33 MeV). The HPGe detector is
placed at $\approx200$ mm from the centre of the crystal and, in
successive measurements, at angles $\theta$ of (12.5, 20.0,
35.0)$^\circ$ with respect to the axis defined by the NaI(Tl)
detector and the CdWO$_4$ crystal. In this way, the overall
angular range covered by the HPGe extends, with variable
efficiency, from $0.5^\circ$ to $44^\circ$, corresponding to
energies of the scattered $\gamma$ rays from 510 keV to 395 keV
and a complementary energy range of Compton electrons inside the
CdWO$_4$ from 1 keV to 116 keV.

It is worth to stress that with this set-up the problem of the
time mark associated to even very low energy events in the crystal
is overcome, because the event trigger is obtained from the
coincidence of 511 keV full-energy signals detected by the NaI(Tl)
and signals detected by the HPGe in the energy range (390 -- 511)
keV. A drawback of this method concerns the intrinsic energy
spread of the 511 keV $\gamma$ rays, which brings the FWHM of the
511 keV line to $\approx 3.0$ keV (the intrinsic HPGe resolution
at this energy is $\approx 1.7$ keV).

For monitoring gain and baseline of the HPGe, a $^{133}$Ba source
is placed near the HPGe detector and shielded by a 25 mm lead
shield from the CdWO$_4$ crystal. For the same purpose, a source
of $^{241}$Am (emitting 60 keV photons) is
 placed near the  CdWO$_4$. Lead
shields have been also introduced which, to some extent, prevent
511 keV (and to a lesser extent 1274.5 keV)  $\gamma$ rays from
 the $^{22}$Na source, from
directly reaching the HPGe detector.

At the arrival of an event trigger, the linear signal from HPGe is
peak analysed (by a 100 MHz, 8k ADC) and recorded, together with
the information on CdWO$_4$ ``signal shape'', on a dedicated PC.
As an example,  the relevant part (470 keV -- 520 keV) of the HPGe
spectrum corresponding to $\gamma$ rays scattered from the
CdWO$_4$ crystal in a 4-days run 
(corresponding to about one tenth of the total statistics) 
at a mean angle $\theta=
12.5^\circ$, is shown in Fig. 4$a$. The broad structure (''bump'')
in the $\gamma$ spectrum is due to 511 keV $\gamma$s Compton
scattered in the crystal, which are in prompt coincidence with the
event trigger. The out-of-coincidence spectrum (Fig. 4$a$)
clearly shows a 511 keV  full energy peak, which is followed
(because of multiple scattering) by a small tail extending from
the full-energy peak down to the Compton edge. Obviously, a
distribution of the same relative size is present at the low
energy side of each channel of the bump and the combined effect of
all channels produces the low-energy tail extending well below the
minimum energy of the scattered $\gamma$s. This effect certainly
contributes, with increasing importance, to the lower-energy part
of the bump. Owing to this reason, only the upper part of the bump
has been considered in the analysis.

In  Fig. 4$b$, the corresponding digitizer pulse amplitude spectrum of
electrons in prompt coincidence is reported. The nominal energy
scale (``apparent electron energy'' $\widetilde{E}_e$) is
normalised to the full-energy peak of 511 keV, corresponding to a
full absorption of the $\gamma$ inside the crystal. Due to the 
non-linear response, the energy scale does not reproduce the true
energy of low-energy electrons.

\nopagebreak
\begin{figure}[h]
\begin{center}
\mbox{\epsfig{figure=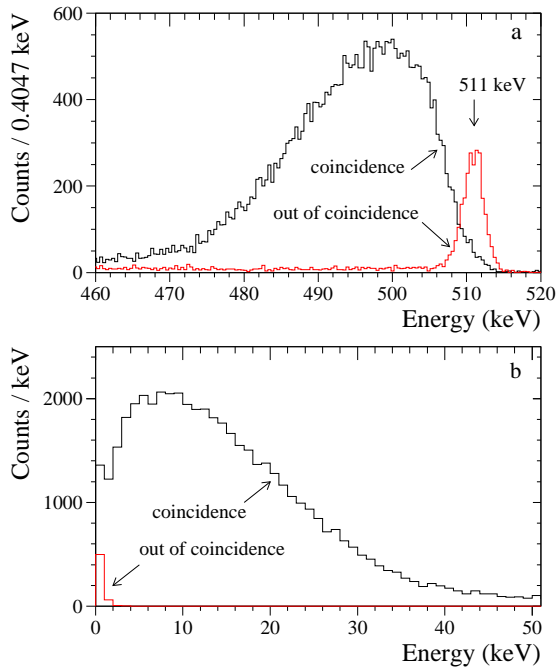,height=9.0cm}} 
\caption{(Color online)  
$a$: Spectrum of the 511 keV $\gamma$ rays scattered by the 
CdWO$_4$ crystal, acquired by HPGe in prompt coincidence with the 
corresponding 511 keV $\gamma$ rays, detected by the NaI(Tl) crystal.
The two spectra are normalised to equal time intervals;
the latter is then multiplied by 20. 
$b$: Distribution of pulse
amplitudes from CdWO$_4$ converted to ``apparent electron
energy'' in keV triggered by the prompt coincidence  between HPGe
and NaI(Tl); and out of coincidence normalised to equal time
intervals. All the spectra refer to mean angle $\theta=
12.5^\circ$ (see Fig. 3).}
\end{center}
\end{figure}

The small rise of the electron spectrum at the lowest energies is
completely accounted for by the background distribution (see the
out-of-coincidence spectrum in 
Fig. 4$b$).

Within the statistical uncertainties, the energy of electrons
associated to a $\gamma$ ray of energy $E_\gamma$ is $E_e=511$
keV -- $E_\gamma$. The bump in the distribution of $E_\gamma$, 
shown in Fig.~4$a$, 
as well as those at $\theta=20^\circ$ and $35^\circ$,
have been divided in bins 2 or 4 keV wide, and for
each bin the distribution of electron signal amplitudes from
CdWO$_4$ has been obtained. Examples of the amplitude spectra of
electrons are presented in Fig. 5 for mean energies 5.1 keV  and
19.1 keV  (from 2-keV bins), and 92 keV (4-keV bin).
Also in this figure, the electron pulse
amplitude has been converted in keV to give the ``apparent
electron energy'' $\widetilde{E}_e$. For each amplitude
distribution, two values of the mean energy are given in the
figure: the ~true mean energy of the scattered electrons~
$\left<E_e\right>$ = 511 keV -- $\left<E_\gamma\right>$, coming from
the mean value $\left<E_\gamma\right>$ of the $\gamma$ energy of
the selected bin on the $\gamma$ bump, and the mean of the
~apparent electron energy~ $\left<\widetilde{E}_e\right>$.

\nopagebreak
\begin{figure}[h]
\begin{center}
\mbox{\epsfig{figure=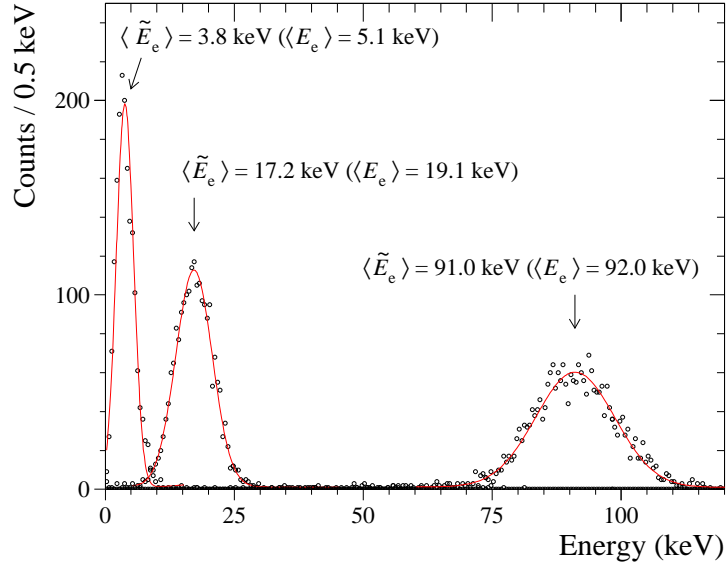,height=8.0cm}} \caption{ (Color
online) Energy spectra of electrons with energies 5.1, 19.1 and
92.0 keV measured by CdWO$_4$ crystal scintillator. The solid lines represent 
the fit to the distributions by a gaussian function.}
\end{center}
\end{figure}

Distributions, as those shown in Fig. 5, have been used for two
purposes:\\
1) from each electron distribution the value of FWHM has been 
extracted and compared with those obtained by irradiating the CdWO$_4$ 
crystal with $\gamma$ rays of similar energies from radioactive sources;\\
2) for each bin of the selected regions of the three $\gamma$ bumps 
(at $\theta=12.5^\circ$, $20^\circ$, $35^\circ$) 
the relative light yield $\left<\widetilde{E}_e/E_e\right>$
of the corresponding electron energy 
distribution has been evaluated.

\subsection{Response of CdWO$_4$ crystal scintillator to low energy
electrons}

The dependence of the energy resolution on energy of electrons and
$\gamma$s from radioactive sources in CdWO$_4$ is reported as a
function of energy in  Fig. 6. The main findings are that no
appreciable differences between electrons and $\gamma$ rays are
apparent in the overlapping energy region. The continuous line
shows a minimum-$\chi^2$ fit of the data with the function
FWHM(\%)$=\sqrt{a/E}$; ($a= 38200\pm400$ keV, $\chi^2$/n.d.f. =
0.45, where n.d.f. is the number of degrees of freedom).

\nopagebreak
\begin{figure}[h]
\begin{center}
\mbox{\epsfig{figure=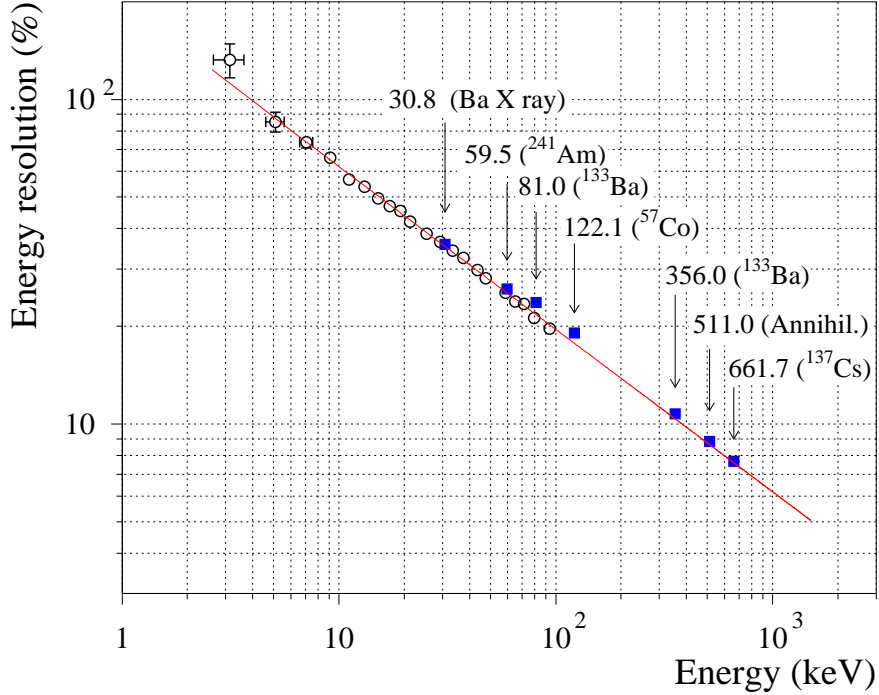,height=10.0cm}} \caption{ (Color
online) FWHM of the distributions of pulse amplitudes from the
CdWO$_4$ counter for Compton electrons and $\gamma$s from various
radioactive sources.}
\end{center}
\end{figure}

Concerning the 
relative light yield,
examples of distributions of the
ratios $\widetilde{E}_e/E_e$, for the same  energy bins of Fig. 5,
are given in Fig. 7. From  this type of distributions, and for
each 4-keV bin in the HPGe bumps at angles $\theta$ of (12.5,
20.0, 35.0)$^\circ$, the 
relative light yield
is derived as
$R=\left<\widetilde{E}_e/E_e\right>$. They are presented as a
function of the true electron energy in Fig. 8. Error bars include
the statistical uncertainties as well as estimated systematic
errors related to the measurement itself and to the procedure  of
data analysis. The 
relative light yield
remains substantially stable
at the 100\% value above 80 keV, decreases slowly at lower
energies down to 95\% at 20 keV and more deeply below this energy.

\nopagebreak
\begin{figure}[h]
\begin{center}
\mbox{\epsfig{figure=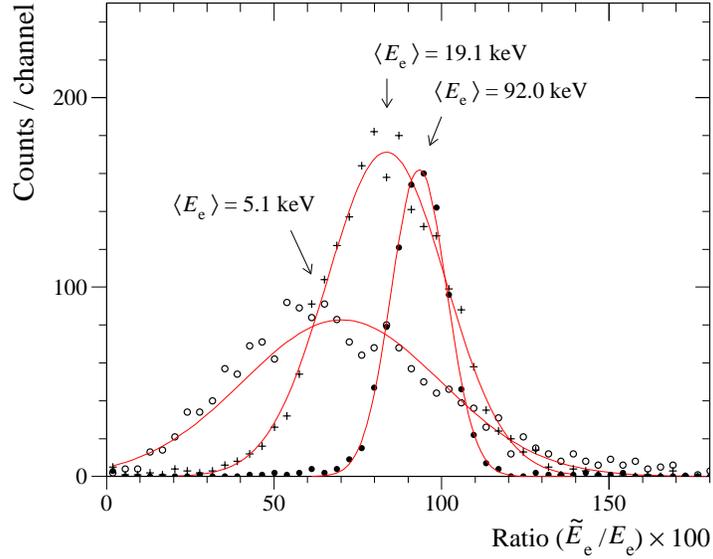,height=8.0cm}} 
\caption{(Color online)  
Examples of distributions of ratio $\widetilde{E}_e/E_e$ 
of the CdWO$_4$ scintillation response
together with their fits by a gaussian function.}
\end{center}
\end{figure}

In CdWO$_4$, no indication is found of an increase of the
luminescence efficiency with decreasing electron energy, as
reported for NaI(Tl). To our knowledge, no previous measurements
with internal electrons have been reported for CdWO$_4$. Compared
with Fig. 10 of \cite{Dore95}, our Fig. 8 shows a slower decrease
of the 
$R$
for energies below 80 keV. We must remark, however,
that the data of \cite{Dore95} are obtained by excitation with
$\gamma$ rays, and show in fact the well known anomalies at the K
edge.

\nopagebreak
\begin{figure}[h]
\begin{center}
\mbox{\epsfig{figure=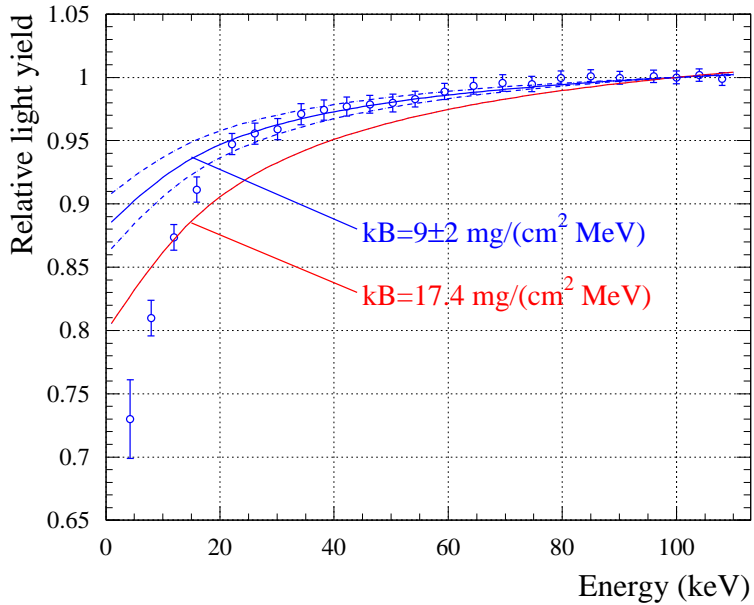,height=8.0cm}} 
\caption{
Relative light yield (normalised to 1 at 100 keV)
as a function of the electron energy. Lines show 
calculations with Eq.~(4) with different $kB$ values.}
\end{center}
\end{figure}

We also calculate here the relative light yield as 
\begin{equation}
R=L_e(E)/E
\end{equation}
with $L_e$ obtained with Eq.~(3) and the Birks factor 
$kB=17.4$ mg~cm$^{-2}$~MeV$^{-1}$, 
the same as was used to describe the experimental data for ions (Fig.~1);
$R$ is normalised to 1 at 100 keV. 
However, as one can see from Fig.~8, the experimental data for electrons
are better described with the value $kB=9.0$ mg~cm$^{-2}$~MeV$^{-1}$, 
when the points at $E>20$ keV are perfectly described, while disagreement
still is present at lower energies.
One could expect that introduction of $kB$ dependent on energy (see \cite{Biz09,Li11,Wil11})
or introduction of additional terms in the Birks equation (see e.g. \cite{Cho52}) would 
allow to improve the description of the data. 

\section{Discussion and conclusions}

(1) It should be stressed, the non-proportionality in the
scintillation response is an important characteristic to estimate the
energy resolution achievable with scintillation crystals \cite{Dore95}.
In accordance with current understanding (see recent reviews 
\cite{Pay09,Mos10,Pay11}), non-proportionality 
of the relative light yield for electrons and $\gamma$ quanta at
low energies is one of the main reasons of poor energy resolution
of a scintillator for $\gamma$s even at higher energies because of
a high probability for a $\gamma$ quantum to interact with a detector
more than once and to lose the total energy through creation of few
lower energy electrons. Many scintillating materials were studied
till now (see \cite{Pay09,Mos10,Pay11,Kho12,Mosz03,Mosz05b,Biza09} and
refs. therein), and the best energy resolution (better that 3\%) was
reached with scintillators which show good proportionality of the light
yield down to $\simeq10$ keV (f.e. LaCl$_3$(Ce), LaBr$_3$(Ce), 
SrI$_2$(Eu) \cite{Pay11}). 
While response of CdWO$_4$ to low energy $\gamma$ quanta was already 
studied \cite{Dore95,Mosz05b}, results for low energy electrons
(down to 5 keV) are presented here for the first time.

(2) Measurements with the accelerator for protons, Li, C, O and Ti 
ions with energies in 1 -- 10 MeV interval, together with the 5.5 MeV 
$\alpha$ particles from $^{241}$Am, allow to obtain data on quenching 
factors interesting for the dark matter studies. 
Description of these data with Eq.~(1)--(3) with the
same value of the Birks factor $kB=17.4$ mg~cm$^{-2}$~MeV$^{-1}$ 
(obtained by fitting the data for protons),
while being not perfect, nevertheless shows relevant agreement,
with the biggest deviation of $\simeq30\%$ for Ti ions.
This further supports the hypothesis \cite{Tret10} that quenching factors
for all ions are not independent and could be described with the same
$kB$ value, if the data are collected in the same experimental conditions and 
are treated in the same way. 
This approach allows to obtain QFs for low energy nuclear recoils on the basis f.e.
of QFs values for a few MeV $\alpha$ particles from internal contamination 
of a detector. Sometimes the QFs for low energy nuclear recoils are known 
with uncertainties much higher than 30\%, 
and the description with Eq.~(1)--(3)
could provide an important estimation of the needed QFs.

(3) It is possible to describe the obtained data on the relative light 
yield for low energy electrons in the CdWO$_4$ by Eq.~(4) 
in a perfect way for $E>20$ keV, however still with disagreement at 
lower energies. Similar description on the basis of the Birks
equation was obtained previously for CaWO$_4$ in \cite{Lan09} and 
for liquid scintillators in the Double Chooz experiment in \cite{Abe11}.
However, for CdWO$_4$ the value $kB=9.0$ mg~cm$^{-2}$~MeV$^{-1}$ 
determined for such a description is different from that obtained
from fit of the data for the ions: $kB=17.4$ mg~cm$^{-2}$~MeV$^{-1}$. 
This unfortunately closes the additional way to obtain QFs for low energy 
nuclear recoils relevant for the DM searches from the light yield 
non-proportionality for the electrons measured with the same scintillator.
This conclusion is not unexpected; it is supported also by the following considerations.
As it is known (see f.e. \cite{Pay11}), for some scintillating materials
(NaI(Tl), CsI(Tl)) the relative light yield for $\gamma$ quanta and electrons
is $R>1$ at low energies.
At the same time, Eq.~(4) can describe only quenched values $R<1$
(as those in Fig.~8 or in \cite{Lan09,Abe11}) but not the 
enhanced $R>1$ (giving at most only $R=1$ 
with $kB=0$), thus it is not suitable for NaI(Tl). 
One could assume that the mechanisms of quenching for ions 
and non-linear response to low energy electrons ($\gamma$s) are different.
Nevertheless, description
for ions by Eq.~(1)--(3) is valid also for NaI(Tl); see f.e. Fig.~13$a$ in 
\cite{Tret10} where the $kB$ value obtained by fitting data for Na recoils
allowed to perfectly describe QFs for I recoils.


\section*{Acknowledgments}

The authors are indebted to S.~Ciattini and F.~Loglio for their
kind collaboration in the measurements of the structural
properties of the CdWO$_4$ crystal. The work of F.A.~Danevich and
V.I.~Tretyak was supported in part by the 
Space Research Program of the National
Academy of Sciences of Ukraine.
We would like to thank the referee for useful suggestions.

\end{document}